\documentclass[final,5p,times]{elsarticle}

\usepackage{amssymb}
\usepackage{hyperref}
\hypersetup{colorlinks=true, linkcolor=blue, citecolor=blue, urlcolor=blue,}
\usepackage{caption}

\journal{NIMB}

\begin{document}

\begin{frontmatter}



\title{A method to detect the VUV photons from cooled $^{229}$Th:CaF$_2$ crystals}

\author[1]{Ming Guan\corref{cor1}}
\ead{guanming-s@s.okayama-u.ac.jp}

\author[2]{Michael Bartokos}
\author[2]{Kjeld Beeks}
\author[1]{Yuta Fukunaga}
\author[1]{Takahiro Hiraki}
\author[1]{Takahiko Masuda}
\author[1]{Yuki Miyamoto}
\author[1]{Ryoichiro Ogake}
\author[1]{Koichi Okai}
\author[1]{Noboru Sasao}
\author[2]{Fabian Schaden}
\author[2]{Thorsten Schumm}
\author[1]{Kotaro Shimizu}
\author[1]{Sayuri Takatori}
\author[1]{Akihiro Yoshimi}
\author[1]{Koji Yoshimura}

\cortext[cor1]{Corresponding author}
\address[1]{Research Institute for Interdisciplinary Science, Okayama University, Okayama, 700-8530, Japan}
\address[2]{Institute for Atomic and Subatomic Physics, Atominstitut, TU Wien, Vienna 1020, Austria}


\begin{abstract}
Thorium-229, with its exceptionally low-energy nuclear excited state, is the only candidate for developing a nuclear clock. $^{229}$Th-doped CaF$_2$ crystals, benefiting from calcium fluoride's wide band gap, show great promise as solid-state nuclear clock materials. These crystals are excited by vacuum ultraviolet (VUV) lasers, which over time cause radiation damage. Cooling the crystals can mitigate this damage but introduces a challenge: photoabsorption. This occurs when residual gas molecules condense on the crystal surface, absorbing VUV photons and deteriorating detection efficiency. To solve this, we developed a cooling technique using a copper shield to surround the crystal, acting as a cold trap. This prevents ice-layer formation, even at temperatures below $-100\,^\circ$C, preserving high VUV photon detection efficiency. Our study detailed the experimental cooling setup and demonstrated the effectiveness of the copper shield in maintaining crystal performance, a critical improvement for future solid-state nuclear clocks operating at cryogenic temperatures.
\end{abstract}


\end{frontmatter}



\section{Introduction}
Thorium-229 has the lowest nuclear-excited energy level of around 8.4\,eV, and it is referred to as an isomeric state    \cite{kroger1976features, Beck2007, vdWense2016nature, Sikorsky2020}. The photons emitted from the radiative decay of this isomeric state are around 148 nm, which is in the vacuum ultraviolet (VUV) region \cite{kraemer2023observation}. This low energy level has led to the concept of developing an ultra-precise nuclear clock based on the transition between the isomeric state and the ground state   \cite{peik2003nuclear, campbell2012single}. A CaF$_2$ crystal, with its wide band gap of 12\,eV   \cite{rubloff1972far,letz2009temperature}, is a promising matrix material for building a solid-state nuclear clock  \cite{rellergert2010constraining, kazakov2012performance}. Recent research has seen significant progress in this area, with $^{229}$Th being successfully excited to the isomeric state either directly by 8.4\,eV lasers  \cite{tiedau2024laser, Elwell:2024qyh, zhang2024frequency} or indirectly by 29\,keV X-rays   \cite{hiraki2024controlling} when thorium ions are doped in CaF$_2$.

In the experiments using CaF$_2$ as a platform to hold $^{229}$Th$^{4+}$ ions, the doped crystals are irradiated with X-ray or laser for the $^{229}$Th isomer production. During the isomer production process, the excitation light source will induce permanent damage in the crystal, which decreases the transmission of signal photons \cite{Beeks:2022hkj}. Furthermore, these damages may also cause degradation of the performance of the solid-state nuclear clock in the future  \cite{kazakov2012performance}. 

One approach to prevent such degradation is cooling the crystals to cryogenic temperatures  \cite{Beeks:2022hkj}, which has also been suggested to improve the accuracy of solid-state clocks in the future  \cite{kazakov2012performance}. Several experiments have already been conducted using $^{229}$Th:CaF$_2$ crystals at low temperatures  \cite{tiedau2024laser, higgins2024temperature}. However, in cryogenic experiments (below $-100\,^\circ$C), a phenomenon has been observed where the detection efficiency for VUV photons from the target crystal decreases with cooling time. This effect is hypothesized to be caused by residual molecules in the vacuum vessel condensing on the cold crystal surface. As these molecules accumulate, they form a layer that absorbs VUV photons. With increasing cooling time, the layer grows in thickness, leading to stronger VUV photon absorption, until the signal photons become undetectable. This phenomenon, referred here to as ice-layer formation or ice-layer absorption, was reported in the supplementary material of   \cite{tiedau2024laser}.
 
Given the advantages of a cooled $^{229}$Th:CaF$_2$ crystal for detecting isomer signals and enhancing the performance of future solid-state nuclear clocks, addressing the issue of the ice-layer formation is crucial. In this report, we introduce a method developed to solve this problem, which can efficiently cool the crystal while preventing ice-layer formation. 

\begin{figure*}[ht]
    \centering
    \includegraphics[width=\linewidth]{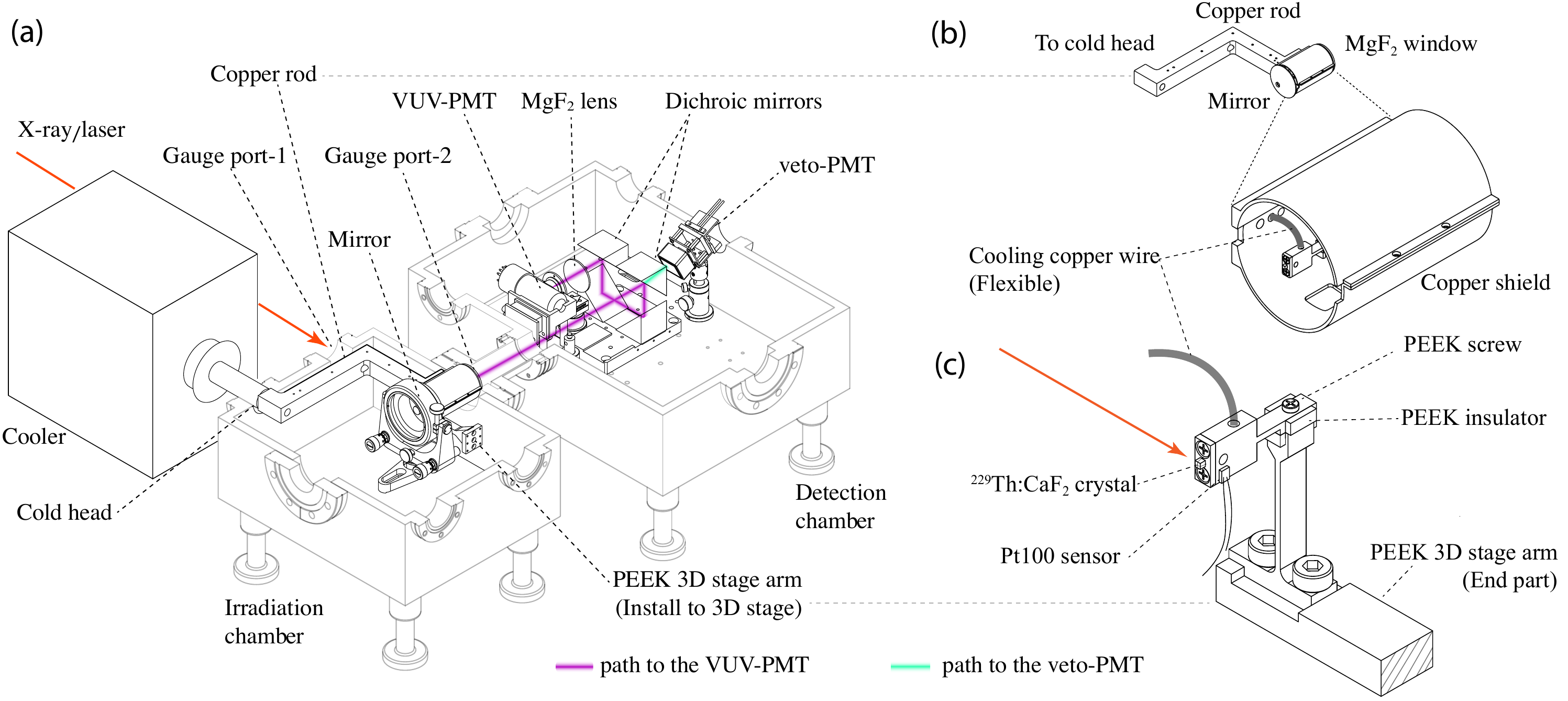}
    \caption{(a) The setup for crystal cooling, isomer production, and detection. (b) The cryogenic conducting copper rod, and the ice-layer suppressing copper shield. (c) The cooling enabled crystal target holder. }
    \label{setup}
\end{figure*}

\section{Method}
\subsection{VUV photon detection setup}
Our experimental setup  \cite{hiraki2024controlling, hiraki2024experimental} was developed to produce the $^{229}$Th isomer states by using 29\,keV X-ray beam and detect the signal photons, at the time when direct spectroscopy by VUV laser was unpromising  \cite{jeet2015results, yamaguchi2015experimental}, due to the large energy uncertainty of the isomer state at that time. This report will introduce the updated setup for cooling the target crystal.

\begin{figure}[b!]
    \centering
    \includegraphics[width=0.95\linewidth]{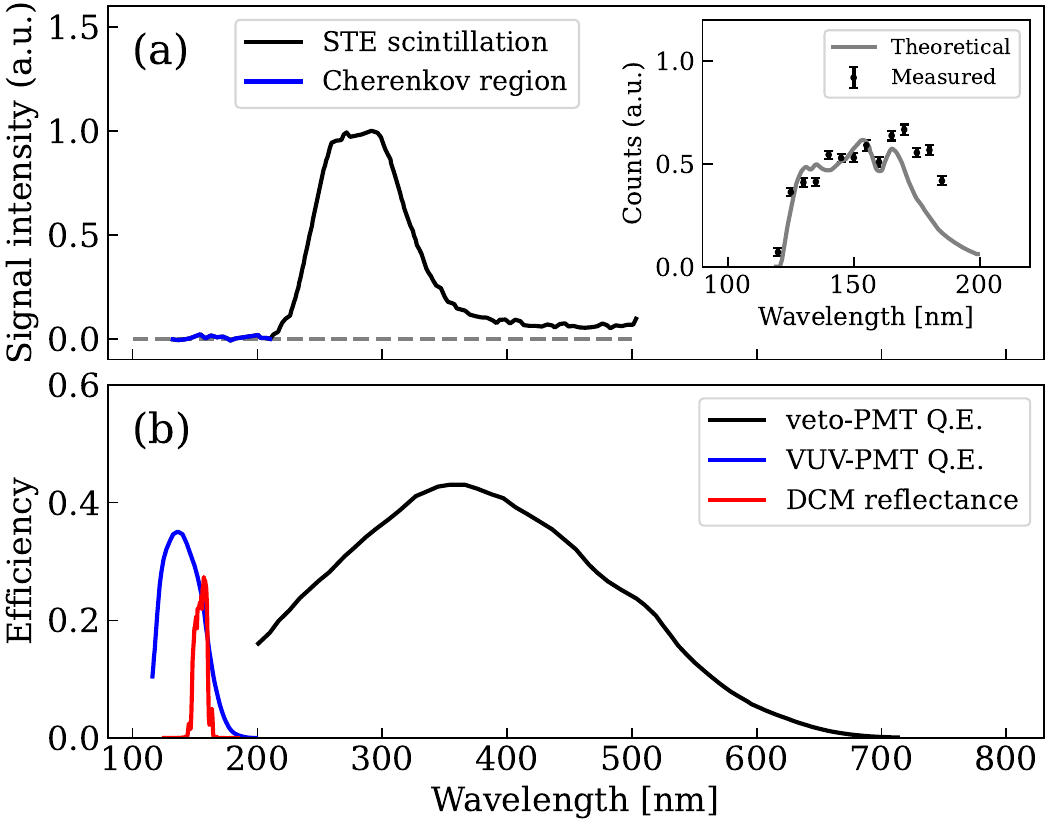}
    \caption{(a) The radioluminescence spectrum of the $^{229}$Th:CaF$_2$ crystal measured by TU Wien group in 2015, where the distinct STE band was observed but the signal counts in Cherenkov region is barely discernible from zero  \cite{stellmer2015radioluminescence}. The inset plot shows the measurement and calculation result for the Cherenkov region, as reported in their recent 2023 study \cite{beeks2023growth}. The (a.u.) on the both vertical axes stands for arbitrary units. (b) Efficiencies of optical devices used in this measurement. Q.E.: quantum efficiency. Q.E. data were taken from Hamamatsu Photonics K. K.  \cite{hamamatsu2007photomultiplier}. DCM reflectance is the total reflectance of 4 dichroic mirrors, measured by our group.}
    \label{efficiency}
\end{figure}

The main structure of the updated setup for cooling the crystal, producing and detecting the $^{229}$Th isomer state, is shown in Fig.~\ref{setup}\,(a). To avoid X-ray induced damage to detectors, we built two separate chambers. We called one of them the ``irradiation chamber'' which was designed for isomer production, and the other one the  ``detection chamber'' for the detection of VUV photons emitted from the radiative decay of the $^{229}$Th isomeric states. The thorium isomers inside the target crystal are populated in the irradiation chamber via the method introduced in   \cite{yoshimi2018nuclear,masuda2019nature}. After the production of the isomer states, the fluorescence photons from the target crystal will be reflected into the detection chamber for signal measurement.

During the measurement of the isomer signal, there are two types of backgrounds. One is the radioluminescence (RL) which mainly originate from the interaction between the $\alpha$, $\beta$ particles emitted from the decay chain of $^{229}$Th and the CaF$_2$ crystal. The normalized RL spectrum of $^{229}$Th:CaF$_2$ is shown in Fig.~\ref{efficiency}\,(a)   \cite{stellmer2015radioluminescence, stellmer2016feasibility}. As shown in the figure, the RL consists of Cherenkov radiation and self-trapped exciton (STE) scintillation  \cite{beeks2023growth}, which dominates in different wavelength regions. The other type of background is the X-ray induced fluorescence, also called afterglow, which will decay to zero within several minutes after the X-ray beam is stopped  \cite{hiraki2024experimental, rao1971afterglow}. In the detection chamber, the dichroic mirrors (DCM) are used to filter out most of the background photons, see the violet light path in Fig.~\ref{setup}\,(a) and red reflectance curve in Fig.~\ref{efficiency}\,(b). The photons coming out of the DCM assembly will be focused with an MgF$_2$ lens and detected by a solar blind photomultiplier tube (VUV-PMT: Hamamatsu, R10454). Meanwhile, another UV-sensitive PMT (veto-PMT: Hamamatsu, R11265-203) can measure the STE light burst in RL, see black curves in Fig.~\ref{efficiency}, which forms an anti-coincident measurement scheme to reject the RL background. The efficiencies of the main optical devices are shown in Fig.~\ref{efficiency}\,(b) \cite{hamamatsu2007photomultiplier}.
More details about the isomer detection in our setup can be found in   \cite{hiraki2024experimental}.

To study the ice-layer formation and its impacts on isomer detection, we use the RL from the target crystal to assess the signal detection efficiency of our setup. As shown in the inset of Fig.~\ref{efficiency}\,(a), the broad Cherenkov radiation spectrum includes a VUV component, which overlaps with the monochromatic isomer signal  \cite{stellmer2015radioluminescence, stellmer2016feasibility, beeks2023growth}. This VUV component can be selected using the preferential reflective band of the DCM assembly, as shown by the red curve in Fig.~\ref{efficiency}\,(b). Since both the RL and isomer photons are emitted within the crystal, we use the selected Cherenkov photons as a VUV source to evaluate our setup's efficiency for detecting isomer signals. 

In this study, we tested the crystal utilized in the X-ray controlling experiment in \cite{hiraki2024controlling}, which yields the VUV-PMT counting rates of approximately 160\,Hz. Additionally, the crystal applied for this study was cut from the same ingot of the ``X2'' crystal, which had been investigated in \cite{tiedau2024laser, zhang2024frequency, nalikowski2024ab}. More details about the synthesis of thorium-doped crystals can be found in   \cite{Beeks:2022hkj, schreitl2016growth}, and optical characterization in   \cite{stellmer2015radioluminescence, beeks2023growth}.

\subsection{Cooling setup\label{cooling_subsection}}
The cooling of the crystal is achieved by using a compact cryogenic cooler (SunPower, CryoTel GT). As shown in Fig.~\ref{setup}\,(a), the cooler is connected to the irradiation chamber via an NW-50 vacuum flange, with its cold head extending into the chamber. The temperature of the cold head can be set in a range from 42 to 77\,K. In the cooling experiment, oxygen-free copper was used to conduct heat away from the target holder, thereby lowering the crystal temperature. During the development of the cooling configuration, various shapes, lengths, and cross sections of copper conductors—such as wire, plate, bar, and rod—were tested to evaluate the cooling capability. All test designs had the constraint that the crystal needed to remain adjustable using a 3D-movable stage.

The purpose of the cooling experiment is to lower the crystal's temperature to approximately 100\,K, reducing the damage by laser  \cite{tiedau2024laser} or X-ray irradiation. Directly attaching the small crystal to the cold copper head or using a copper holder can achieve this target temperature; however, such methods are not feasible because we need to frequently install/uninstall, and anneal (400\,$^\circ$C)  \cite{hiraki2024experimental} the target crystal, and we need to change the position of the crystal to gain maximum detection efficiency. Instead, we cool the stainless steel target holder to achieve the cooling of the crystal. As illustrated in Fig.~\ref{setup}\,(c), the crystal ($\sim$1\,mm$^3$) is mounted on the front surface of the target holder, which has a hole in its body. A flexible copper wire connects the target holder to a copper rod or bar, enabling positional adjustments. The opposite end of the copper rod or bar is fixed to the cold head.  The insulator and 3D stage arm are made of PEEK plastic to mitigate heat conduction from the stage.

As shown in Fig.~\ref{setup}\,(c), a Pt100 sensor is attached to the target holder to monitor temperatures. The readout temperature values are recorded in a logger (GraphTech, GL840), where we regard this recorded temperature as the temperature at the target.  To control the temperature of the crystal, we attached heaters to the copper conducting structure. One of the heaters is the $10\times20$\,mm$^2$ rectangle heater, Micro-Ceramic Plate Heater (Misumi, MMCPH-20-10). The other one is the round-shaped plate heater with a diameter of 23\,mm (Thorslab, HT19R). Both heaters are 1.3\,mm thick and their installation depends on the experiment's purposes.

\subsection{Vacuum improvement practices}
When we started the cooling experiment, the vacuum condition was 10$^{-2}$\,Pa, as measured with a vacuum gauge (Pfeiffer Vacuum, PKR360) at the gauge port-1 in Fig.~\ref{setup}\,(a), and at that time the detection rate of the VUV-PMT decreased very rapidly in the initial cooling test, due to the ice-layer absorption.  This relatively low vacuum condition was due to several factors: the low pumping speed of the evacuation connection, leakage of the flanges, and contamination-induced outgassing and evaporation within the chamber. This vacuum condition impedes the cooling experiments essential for the studies of $^{229}$Th isomer, necessitating efforts to address this issue. 

A natural approach to suppressing ice-layer formation is to improve the vacuum quality inside the chamber, thereby minimizing the number of residual molecules  \cite{nakhosteen2016handbook}.  To improve the vacuum, two turbomolecular pumps (TMPs) were directly mounted on the chamber lid: a 220\,l/s TMP (Osaka Vacuum, TG220FCAB) and a 250\,l/s TMP (Pfeiffer Vacuum, HiPace 300).  In addition, to suppress leakage and outgassing, the chambers were completely disassembled, cleaned, and baked. After reassembly, each flange was checked for leaks using a helium leak detector (Agilent Tech., VSPD03). Leakage for each flange port was controlled to be below $10^{-11}$\,Pa$\cdot$m$^3$/s. The entire setup was baked at 70\,$^\circ$C to accelerate the outgassing and the emission of molecules from the chamber’s inner walls. A mild baking temperature of 70\,$^\circ$C was chosen to avoid damaging the rubber O-rings. After these procedures, the vacuum level was improved from 10$^{-2}$\,Pa to 10$^{-5}$\,Pa.

\subsection{Improved cooling scheme: copper rod conductor and shield}
To improve the cooling performance and suppress ice-layer absorption, we updated the cooling scheme and developed a copper shield, which is illustrated in Fig.~\ref{setup}\,(a)-(c).  In this improved cooling scheme, the conducting copper rod has a right-angle shape with a cross section of $20\times20$\,mm$^2$, and the lengths of its two sides are 120\,mm and 180\,mm, respectively. The shield is a cylindrical shell with an inner diameter of 39\,mm, a length of 75\,mm, and a thickness of 1\,mm, as shown in Fig.~\ref{setup}\,(b), and it can be mounted on the copper rod. In the cooling test, it was found that the cold shield temperature was usually lower than the crystal temperature by ten degrees.

The shield is designed to absorb residual molecules and prevent them from condensing on the crystal surface.  As shown in Fig.~\ref{setup}\,(b), one end of the shield is closed off with the parabolic mirror with 1\,mm gap kept between the mirror and the copper shield to avoid contact when the copper shrinks at low temperatures. The other end is completed with the  MgF$_2$ window which separates the vacuum between the irradiation chamber and the detection chamber. The parabolic mirror, copper shield, and MgF$_2$ window together enclose the target holder, creating a semi-closed space free from molecular flow. Meanwhile, the cold shield works as a cold trap that captures the molecules inside it, therefore it features high vacuum conditions for the crystal cooling. The vacuum level reaches $5\times10^{-6}$\,Pa during the operation of the cooler, when the gauge was installed to the gauge port-2 in Fig.~1\,(a).
\begin{figure}
\centering
\includegraphics[width=\linewidth]{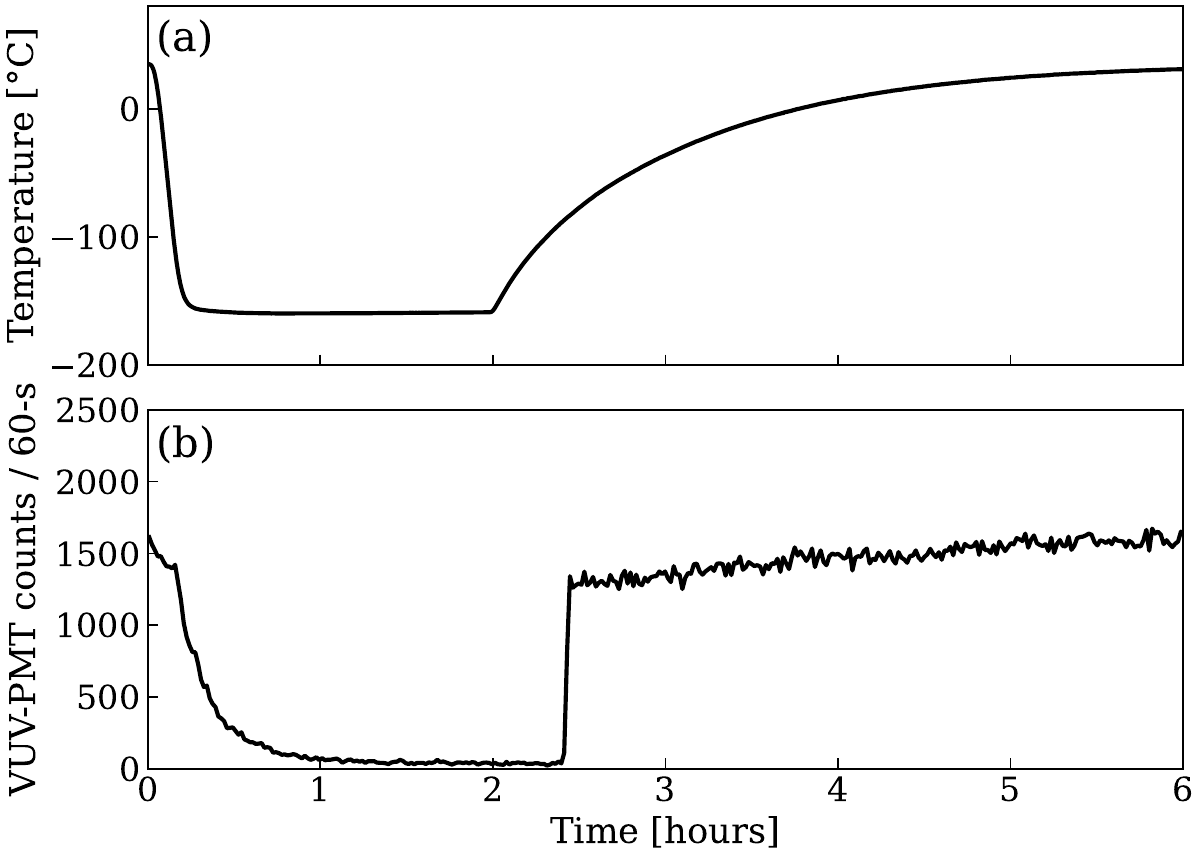}
\caption{(a) Temperature of the crystal when the copper bar is used to conduct heat. (b) The VUV-PMT counting rate during the cooling and warming cycle.}
\label{config_B}
\end{figure}

\section{Experiments and Results}
\subsection{Cooling test}
With various copper conductors, cooling tests were performed to investigate the cooling capacity (lowest temperature reached and temperature adjustment speed) and the ice-layer formation. During the cooling and warming cycle, the VUV-PMT counts were monitored to see how the ice-layer absorption changed with temperature and over time. The results are discussed below.

\subsubsection*{Thin copper bar conduction}

A slim copper bar (cross section $2\times4$\,mm$^2$) was earlier tested for its cooling capability. As shown in Fig.~\ref{config_B}\,(a), the cooler was turned on at  $t=0$\,h and off at $t=2$\,h. Due to the small volumes of this copper bar and target holder, the cooling process was rapid, with the crystal reaching a minimum temperature of $-155\,^\circ$C within 20 minutes in this test. In contrast, the warming up process was relatively slow when the heater was not used. This is likely due to the lack of external heat flow to the target holder.

In panel (b), the VUV-PMT counting rate during the cooling and warming cycle is shown. During the cooling process, the detection rate decreases drastically. Within one hour, the VUV-PMT counting rate approaches zero, with the remaining non-zero values attributed to the dark count rate of the PMT.  In the warming-up session, an observed phenomenon is that when the crystal temperature reached $-100\,^\circ$C, the detector counting rate suddenly increased to a relatively high level. This phenomenon is attributed to the evaporation of the layer at this temperature. 
 
\subsubsection*{Thick copper rod and shield}

\begin{figure}[t!]
    \centering
    \includegraphics[width=\linewidth]{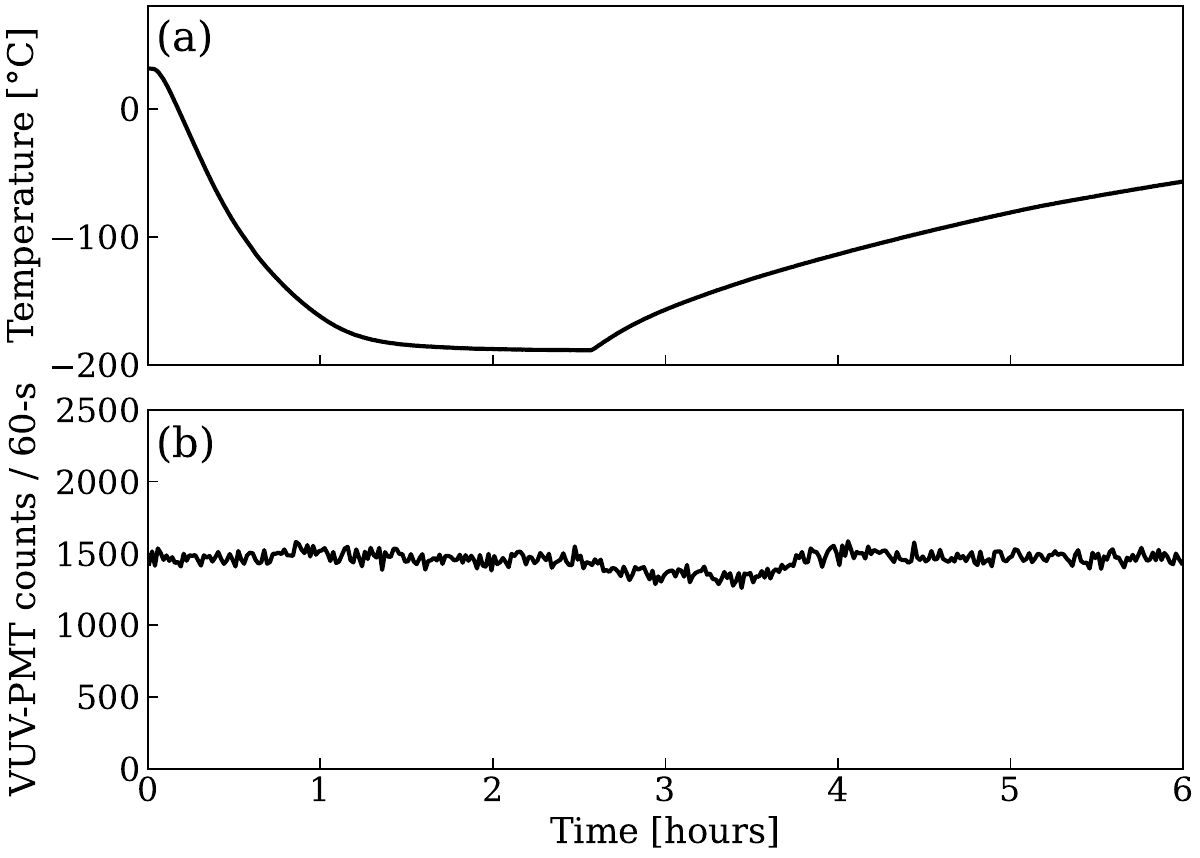}    \caption{(a) Temperature of the crystal, when the copper rod is used to conduct heat, and the shield and the MgF$_2$ window is installed. (b) The VUV-PMT counting rate during the cooling and warming cycle.}
    \label{config_C}
\end{figure}

When the copper rod (cross section $20\times20$ \,mm$^2$) and shield are installed, the crystal's temperature and detector's counting rate during the cooling and warming cycle are shown in (a) and (b) of Fig.~\ref{config_C}, respectively. The copper rod is much thicker than the bar, resulting in longer cooling (2 hours) and warming (8 hours) times for the crystal when the heater power is zero.  On the other hand, compared with the copper bar, the thermal resistance is further reduced, allowing the crystal to reach a lower temperature of $-190\,^\circ$C. 

As seen in panel (b), the introduction of the copper shield and the MgF$_2$ window significantly suppresses the ice-layer formation, as indicated by the stable VUV-PMT counting rate when the crystal is cooled below $-100\,^\circ$C. Therefore, during the full cooling and warming cycle, the VUV-PMT count rate was unaffected by layer absorption. The slight change in the detector counting rate near $t=3$\,h could be attributed to the variation of the detection efficiency caused by multiple factors. For example, the crystal's position could be shifted due to the contraction/expansion of the copper rod during temperature changes, and the molecules trapped in the copper shield would evaporate during the warming-up process, which presents VUV absorptions. The impact of these factors requires further studies.

\subsection{Ice-layer formation in various temperatures}

\begin{figure}[t!]
    \centering
    \includegraphics[width=\linewidth]{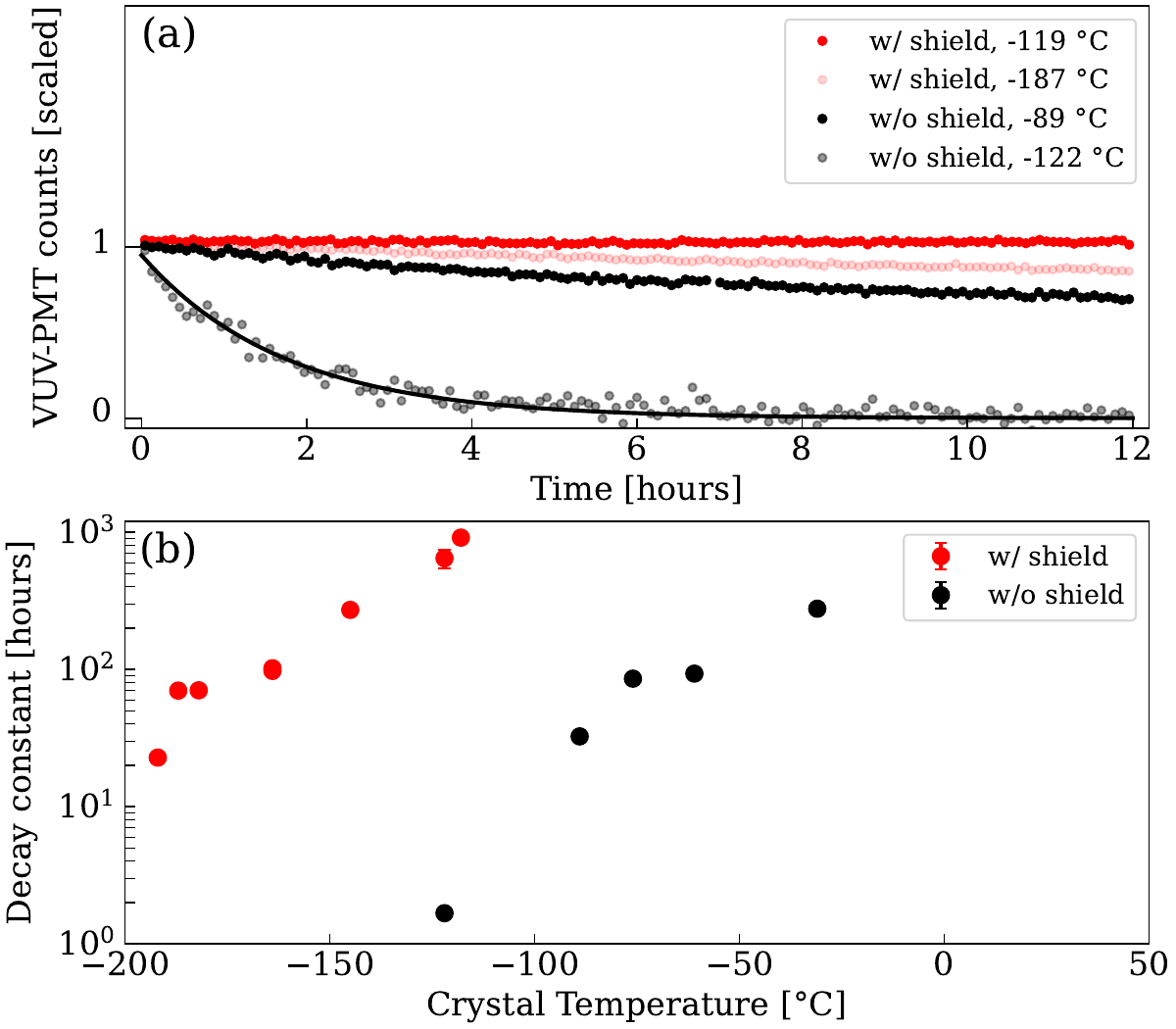}
    \caption{(a) The ice-layer absorption measured at several temperatures with and without the shield. (b) The fitted decay constants with and without the copper shield and the MgF$_2$ window.}
    \label{shield_tau}
\end{figure}
We also studied how the VUV photons were absorbed by ice-layer by monitoring VUV-PMT event rates  while maintaining the crystal at a constant temperature. A series of 12-hour measurements at different temperatures, both with and without the copper shield, were conducted, with four examples presented in Fig.~\ref{shield_tau}\,(a). To compensate the shifts of VUV transmission of the crystal in these long term measurements, the detector counting rates are normalized to one at time zero. As shown in the figure, the counting rate decreases exponentially over time and can be well-fitted with a pure exponential function. For example, the exponential fitting curve for the data point at $-122\,^\circ$C is plotted.

In the lower pad, Fig.~\ref{shield_tau}\,(b) shows the decay constants measured in different temperatures with and without the copper shield and MgF$_2$ window. It is clear that the decay constants become smaller at lower temperatures, which means the signal tends to be absorbed faster at lower temperatures, while the copper shield and the MgF$_2$ window can make a significant difference in the decay constant, see the red and black points. At the same cryogenic temperature, the presence of the copper shield extends the time constants by several orders. For example, when the crystal was stabilized near $-120\,^\circ$C, the decay constant was prolonged from 1.7 hours (without shield) to 647 hours (installed shield). 


\section{Discussion and Conclusion}

From the above results, the following findings were obtained regarding the ice-layer formation and its suppression.

\renewcommand{\labelitemi}{\hspace*{1em}} 
\begin{itemize}

\item
It was observed that the ice-layer formation progresses in proportion to time and that the growth rate depends on the vacuum level and crystal temperature. In particular, even in a high vacuum environment at the 10$^{-5}$\,Pa level, a sudden deterioration in the signal was observed when the temperature dropped below $-100\,^\circ$C. 

\item
To effectively suppress the ice-layer formation,  covering the crystals with a parabolic mirror, a copper shield and an MgF$_2$ window to form a closed space is effective. Furthermore, it was found that cooling this shield and using it as a cold trap can suppress ice-layer formation even below $-100\,^\circ$C.

\item 
Finally, a long time constant of more than 20 hours was obtained when the shield and the MgF$_2$ window was installed even at very low temperatures $-190\,^\circ$C, achieving a long experiment time. Further extension of the time constant can be expected by further improving the vacuum condition or shielding structure.
\end{itemize}

This report presents a method to suppress the ice-layer formation in the $^{229}$Th:CaF$_2$ crystal cooling experiment, which provides a technical concept to maintain isomer detection at very low temperatures. The method is not only effective for low-temperature operation in future solid-state nuclear clocks, but also applicable as a general concept to suppress the ice-layer formation that commonly occurs in VUV optical systems at cryogenic temperatures.

\section*{Acknowledgements} 
{This work was supported by JSPS KAKENHI Grant Numbers JP21H04473, JP23K13125, JP24K00646, JP24H00228, JP24KJ0168. This work was also supported by JSPS Bilateral Joint Research Projects No. 120222003.  This work has been funded by the European Research Council (ERC) under the European Union’s Horizon 2020 research and innovation programme (Grant Agreement No. 856415) and the Austrian Science Fund (FWF) [Grant DOI: 10.55776/F1004, 10.55776/J4834, 10.55776/ PIN9526523]. }




\bibliographystyle{elsarticle-num}
\bibliography{cooling_literatures}

\end{document}